
\input phyzzx.tex

\def\quadg {g^{ab}\nabla_a\nabla_b}
\def\quadgg {\hat g^{ab}\nabla_a\nabla_b}

\def\qgg {\Delta_{\hat g}}
\def\d {\partial}
\def\v {\varphi}
\def\l {\lambda}
\def\a {\alpha}
\def\g {\hat g}
\def\z {\bar z}
\def\t {\theta}
\def\f {\phi}
\def\b {\beta}
\def\G {\Gamma}
\def\F {\Phi}
\def\picc {\scriptscriptstyle}
\def\s {\sigma}
\def\e {\epsilon}
\def\D {\Delta}
\def\dd {\delta}
\def\L {\Lambda}

\titlepage
\title{A continuum approach to 2D-Quantum Gravity for c$>$1}

\author {M. Martellini\foot{On leave of absence from Dipartimento di Fisica,
Universit\'a
di Milano, Milano, Italy and I.N.F.N., Sezione di Pavia, Italy}}
\address{I.N.F.N., Sezione di Roma "La Sapienza", Roma, Italy}
\author {M. Spreafico}
\address{Dipartimento di Fisica, Universit\`a di Milano, Milano, Italy
and I.N.F.N., Sezione di Milano, Italy}
\author {K. Yoshida}
\address{Dipartimento di Fisica, Universit\`a di Roma, Roma, Italy and
I. N. F. N., Sezione di Roma, Italy}
\abstract{Two dimensional induced quantum gravity with matter central charge
$c>1$ is studied taking a careful consideration of both diffeomorphism and
Weyl symmetries . It is shown that, for the gauge fixing condition $R(g)$
(scalar curvature)=$const$, one obtains a modification of the
David-Distler-Kawai version of KPZ scaling. We obtain a class of models
with the real string tension for all values $c>1$. They contain an
indeterminate parameter which is, however, strongly constrained by the
requirement of non triviality of such a model. The possible physical
significance of the new model is discussed. In particular we note that it
describes smooth surfaces imbedded in $d$-dimensional flat space time for
arbitrary $d$, consistently with recent numerical result for $d=3$.}
\endpage

\pagenumber=1

\vglue 0.6cm
{\bf 1. Introduction}
\vglue 0.4cm
Recently we have witnessed the
remarkable progress in understanding the quantized
theory of gravity in 2 space-time dimensions in presence of matter fields
(induced gravity) [1] [2].

One very serious problem still remains. This is the seeming inability to go
beyond the matter
central charge, $c$, greater than one. (or in the case of the free
bosonic matter field, $c=d=$number of species of such bosons).
This difficulty can be seen typically in the by now classical expression [2]
for the string tension, initially given by Knizhnik, Polyakov and
Zamolodchikov [1]
$$
\G(h)={1-h\over 12}\left[ d-25-\sqrt{(25-d)(1-d)}\right]+2
\eqn\pippo
$$
($h$ is the genus of the world surface which is taken to be fixed.)

$\G(h)$ becomes complex for $1<d<25$ and we see no way of improving the above
expression within the approaches advanced so far including the powerful matrix
method [3].

In some sense, such a difficulty is foreseen. The large value of $d$ implies
the strong coupling regime for the gravitational interaction of matters and
essentially semiclassical base underlying the conventional approach breaks
down.
(This is particularly clear in DDK [2])
Indeed one may doubt the basic idea of "summing up" more or less smooth
2-dimensional surfaces to evaluate the partition function for the 2-dimensional
quantum gravity. It is known for long time [4] from the work based on the
numerical simulation that there is the region of quantum fluctuation of
Liouville field (representing the gravitational degree of freedom) which tends
to push the dominant configurations toward the singular (spiky) ones for
$d\geq 1$.

There are several models recently advanced, however, which, although in
certain aspect resemble the 2D gravity in conformal gauge, avoids the
difficulty at $d=1$.
One of the most celebrated of such models is the black-hole (BH) conformal
field theory of E. Witten [5]. As it has been noted in the original work,
one sees no indications of breakdown at $d=1$ (original paper refers to
$d=1$. But this can be generalized by coupling to it $d-1$ "matters"
in somewhat arbitrary fashion) [6].

There are several other instances of such "pseudo 2D gravity" models existent
which share certain common characteristics.
They all seem to avoid the problem at $d=1$, or at least nothing special
happens when $d$ is increased from $0$ to $1$.
At the more technical level, they will contain "Liouville field"
which is generally accompanied by another
scalar field of opposite sign. In case
of Witten's BH conformal field theory such an extra scalar degree of freedom
comes from U(1) gauge field in SL(2R)/U(1) WZNW model.

In this work, we have tried to reproduce the effects just discussed starting
from usual 2D quantum gravity model in conformal gauge.
Our point of departure from the older approach (DDK) is in the interpretation
of the fixed area partition function (in the following it is always intended
to operate with surfaces of fixed genus) [7]
$$
Z({\cal A})= \int "Dg" Z_{\picc{M}}(g) \delta\left(\int d^2 x \sqrt{g} -
{\cal A}\right)
\eqn\due
$$
where $g_{ab}$ is the 2D world sheet metric, $Z_{\picc{M}}(g)$ is
the matter partition function in the background metric $g_{ab}$.

We claim that fixing the area of the world sheet to define $Z({\cal A})$ by eq.
\due\ is essentially equivalent to fix the gauge with respect to the Weyl
symmetry under
$$
g_{ab}\to e^{\s(z,\z)} g_{ab}
\eqn\tre
$$

We know that the symmetry under \tre\ is respected in quantum gravity
along with the symmetry under diffeomorphisms of the world sheet by virtue of
$$
c_{tot}~(total~ central~ charge)=0
$$

Now one way of gauge fixing with respect to the Weyl symmetry is
to fix locally the scalar curvature
$$
R_{g}=const
$$

For instance, for any metric $g$ with genus $h\geq 2$, one can find an unique
metric $g'$ ("slice") such that
\par\noindent
{\it 1)} $g'$ is conformably equivalent to $g$
$$
g=e^{\s}g'
$$
{\it 2)}
$$R_{g'}=-1$$

Thus, for any metric $g_{ab}$ one has an unique Weyl equivalent metric $g'$
such that
$$
\eqalign{
&R_{g'}=const=R_0\cr
&{\cal A}'=\int d^2x \sqrt{g'} = {8 \pi\over R_0}
{1\over 8\pi} \int d^2 x \sqrt{g'}
R_{g'}={8\pi\over R_0}(1-h)~~(fixed)\cr}
$$
where $R_0$ can be normalized to be equal to $-1$ for the case $h\geq2$.

Because of the correspondence between the fixed area (global condition) and
the $R_g$ (local condition), and if we insist to use only local gauge fixing
condition, we propose to modify the conventional path integral representation
of $Z({\cal A})$ by adding to it the corresponding Faddeev Popov (FP)
determinant defined by
$$
N= vol(Weyl) \times Det\left [ {\delta\over \delta\s} (R_{g'=e^{\s}g}+1)
\right]_{\s=0}
$$
where $vol(Weyl)$ is the (divergent) total volume of Weyl group.

Since under
$$
g\to g'=e^{k\s} g
$$
one has
$$
R_g\to R_{g'}=e^{-k\s} (R_g - k \quadg\s)
$$
the FP determinant is the determinant of the operator
$$
\eqalign{
&{\delta\over\delta\s} (R_{g'=e^{\s}g}+1)\cr
=&{\delta\over\delta\s} (e^{-k\s} (R_g - k\quadg\s) +1)\cr
=&-k (R_g +\quadg)\cr}
$$
thus the FP determinant becomes
$$
N=vol(Weyl) \times Det [-k (R_g +\quadg)]
\eqn\auno
$$
($k$ can be chosen later conveniently).

To obtain the local action in the final formula for the partition function, we
introduce a pair of fermion fields $\psi$ and $\bar\psi$ (Weyl ghosts) to
substitute in eq. \auno
$$
Det [-k (R_g +\quadg)]=
\int D_g\psi D_g\bar\psi e^{{1 \over 2}k
\int d^2\xi \sqrt{g}\bar\psi(\xi) (R_g+\quadg)\psi(\xi)}
\eqn\adue
$$

Note that the integral measure $ D_g\psi D_g\bar\psi$ depends
non trivially on the background metric $g$.

To simplify the path integral, we assume again the ansatz due to Distler and
Kawai (DK). By our choice of conformal gauge with respect to the
diffeomorphisms
invariance, we write as in ref. [2]
$$
\g_{ab}= e^{\a\f} \g_{ab}(\tau)
\eqn\atre
$$
where $\g_{ab}(\tau)$ is the "flat" background metric parameterized by the
moduli parameter only. The rescaling $\a$ will be determined by the
consistency conditions. With the choice of gauge eq. \atre, we can apply
DK ansatz which allows us to write for instance
$$
 D_g\psi D_g\bar\psi=e^{S_{\picc L}'[\f,\g]} D_{\g}\psi D_{\g}\bar\psi
\eqn\aquattro
$$
$D_{\g}\psi D_{\g}\bar\psi $ is a translationally invariant measure and
$S_{\picc L}'[\f,\g]$ is the Liouville type local action.
\vglue 0.6cm
{\bf 2. Effective action of generalized Liouville model}
\vglue0.4cm
Inserting our factor $N$, eq. \auno, with eq. \adue\ and \aquattro,
into the path integral expression for $Z({\cal A})$,
we obtain the desired modification
of DDK formula for the fixed area partition function
$$
\eqalign{
Z({\cal A}) = \int &D_{\hat g}X D_{\hat g}\phi
  D_{\hat g}b D_{\hat g}c D_{\hat g}\psi D_{\hat g}\bar\psi d\tau]\cr
& e^{-S_{\picc M}[X;\hat g]-S_{\picc GH}[b,c;\hat g]-S_{\picc L}''[\phi;\hat
g]-
  S_{\picc W}[\psi,\bar\psi;\hat g]} \delta\left(\int d^2 x \sqrt{g} -
{\cal A}\right)\cr}
\eqn\acinque
$$
where
$$
\eqalign{
&X=matter~ field\cr
&\f=Liouville~ field\cr
&\psi, \bar\psi=Weyl~ ghosts\cr
&b,c=diffeomorphism~ ghosts\cr
&S_{\picc M}[X;\hat g]=matter~ action~ (Polyakov's~ action)\cr
&S_{\picc GH}[b,c;\hat g]=diffeomorphism~ ghosts~ action\cr
&S_{\picc L}''[\phi;\hat g]=Liouville~ type~ action~ (coming~ from~ both~ Weyl~
and~diff.~ contribution)\cr
&S_{\picc W}[\psi;\hat g]=Weyl~ determinant\cr}
$$

The principal modification here is in the presence of the term
$S_{\picc W}[\psi,\bar\psi;\hat g]$ containing the Weyl ghosts
$$
S_{\picc W}[\psi,\bar\psi;\hat g]
={1 \over 2}k
\int d^2\xi \sqrt{\hat g}\bar\psi(\xi)[-(R_{\hat g}-
\alpha\quadgg\phi+\quadgg)]\psi(\xi)
\eqn\asei
$$

Together with the Liouville action $S_{\picc L}''$, they give the modified form
of gravitational action
$$
\eqalign{
S_{\picc grav}&=
S_{\picc L}''[\phi;\hat g]+S_{\picc W}[\psi,\bar\psi;\hat g]
={1\over 8\pi}\int d^2\xi\sqrt{\hat g}
(\hat g^{ab}\hat{\nabla}_a\phi\hat{\nabla}_b\phi - Q R_{\hat g}\phi)+\cr
&+{k \over 2}
\int d^2\xi \sqrt{\hat g}\bar\psi [ -(R_{\hat g}-
\alpha\quadgg\phi+\quadgg)]  \psi\cr}
\eqn\asette
$$

The coefficient of the kinematical term for the Liouville
field $\phi$ is normalized by rescaling (cf. eq. \atre).
Writing $q=4\pi k$ and with the Laplacian $\qgg$ defined by
$$
\eqalign{
&q=4\pi k\cr
&\qgg=\quadgg\cr}
$$
eq. \asette\ becomes
$$
S_{\picc grav}={1\over 8\pi}\int d^2\xi \sqrt{\hat g}
[-\phi\qgg\phi -QR_{\hat g}\phi-q\bar\psi\qgg\psi
-q\bar\psi R_{\hat g}\psi+\alpha q\bar\psi\psi\qgg\phi]
$$

Note that the Weyl ghost action  \asei\ has the usual form of a $b-c$
ghost system if one makes the identification
$$
\eqalign{
&b\equiv \d\bar\psi\cr
&c\equiv\psi\cr}
$$
with conformal dimension respectively $1$ and $0$.

The free part of ghost action \asei\ is written in the form
$$
\int b\bar\d c
$$
with correspondent stress energy tensor and ghost number current, given by
$$
\eqalign{
&T=-b\d c\cr
&J=bc \cr}
$$

Following ref. [8], one can give the equivalent bosonic system with
$$
\eqalign{
&T_{\picc B}= -{1\over 2} [(\d\v)^2+ Q \d^2\v]\cr
&J_{\picc B}=i \d\v\cr}
$$
with new bosonic field $\v(z,\z)$ satisfying
$$
\v(z)\v(w)\sim -ln |z-w|^2
$$

The algebra between $T_{\picc B}$ and $J_{\picc B}$ is identical with that
of $T$ and $J$\footnote{*}{We acknowledge H. Kawai on the illuminating
discussion on this point.}.
The relationship between the bosonic field $\v$ and its "solitonic" field
$\psi$ is
$$
\psi\sim e^{i \v}
$$

Treating perturbatively the rest of Weyl action, we obtain the equivalent
bosonic form of \asei
$$
S_{\picc W}=
{q\over 8\pi}\int \sqrt{\hat g} [-\v\d\bar\d\v-i(1+A)\v R_{\hat g}+
2i{\alpha A\over 2}\v\d\bar\d\phi ]
\eqn\bquattro
$$
where the terms multiplied by the new renormalization constant $A$ correspond
to the interaction part of \asei. In fact, due to the interaction, one has to
make the identification
$$
\d\bar\psi \psi = i A \d\psi
$$
and thus
$$
\eqalign{
&- q \bar\psi R_{\hat g}\psi+\a q\bar\psi\psi\qgg\phi\cr
&\equiv -{q\over \d} (\d\bar\psi)\psi R_{\hat g} + \a {q\over \d} (\d \bar\psi)
\psi\d\bar\d\phi\cr}
$$

Writing
$$
\eqalign{
&\tilde Q \equiv (1+A)\sqrt{q}\cr
&B\equiv \a {A\over 2}\sqrt{q}\cr
&\v\to\sqrt{q}\v\cr}
$$
\bquattro\ takes the form
$$
\int \sqrt{\g}[-\v\qgg\v+2iB\v\qgg\phi-i\tilde Q\v R_{\g}]
$$

After "Wick rotating" $\v$ field to avoid the imaginary coupling constant,
one obtains
$$
S_{\picc W}={1\over 8\pi}\int \sqrt{\g}[\v\qgg\v-2B\v\qgg\phi+\tilde Q\v
R_{\g}]
$$

Finally, we have our expression of gravitational action
$$
\eqalign{
&S_{\picc grav} = S_{\picc L}'' + S_{\picc W}=\cr
&={1\over 8\pi} \int d^2\xi \sqrt{\g} [-M_{i,j}\Phi^i\qgg\Phi^j-Q_i\Phi^i
R_{\g}
   ]\cr}
\eqn\botto
$$
where
$$
\eqalign{
&\Phi^i\equiv(\Phi^1,\Phi^2)=(\phi,\v)\cr
&Q_i\equiv(Q_1,Q_2)=(Q,-\tilde Q)\cr}
$$
with
$$
\eqalign{
&M_{i,j}\equiv\left(\matrix{1&B\cr
B&-1\cr}\right)\cr}
$$

There is an apparent similarity between our action \botto\ and the action
proposed in ref. [9]. One must realize, however, the origin as well as the
interpretation of form \botto, are completely different. In particular,
the extra scalar $\v$ is introduced in ref. [9] as some form of bosonization of
usual diffeomorphisms ghost ($c$ with weight $-1$), while we are
introducing such an extra scalar as the bosonized form of a new Weyl ghost.
Naturally this difference has an immediate consequence in the respective
counting of the central charge, as we will see below.
\vglue0.6cm
{\bf 3. Constraints on the model}
\vglue0.4cm
The stress energy tensor corresponding to $S_{\picc grav}[\f,\v]$, eq. \botto,
is
$$
T_{\picc grav} = -{1\over 2}[M_{i,j}\d\Phi^i\d\Phi^j+Q_i(\d^2)\Phi^i]
\eqn\bnove
$$

The gravitational contribution to the central charge in our model is then
$$
c_{\picc grav}=2+3M^{i,j}Q_iQ_j
\eqn\c
$$
where $M^{i,j}$ is the inverse matrix of $M_{i,j}$.

With respect to the stress-energy tensor \bnove, the area operator
$\sqrt{\g} e^{\a\f}$ must have the conformal dimension $1$. So that
$$
\Delta(e^{\a\f})=
-{1\over 2} \alpha_1 M^{1,j}(\alpha_j+Q_j)=1
\eqn\cuno
$$

Further, the condition of conformal invariance gives
$$
c_{\picc tot}=c_{\picc grav}+d-26=0
$$

Substituting \c\ for $c_{\picc grav}$, one obtains the equation for $Q_i$
$$
Q_1^2+2BQ_1Q_2-Q^2_2+{1+B^2\over 3}(D-24)=0
\eqn\ctre
$$

For the moment, we have no other obvious constraints to go with.
We now look for the reasonable solutions satisfying eq. \cuno\ and \ctre.
Only, we "phenomenologically" impose the condition that such a solution should
reduce to the DDK solution for $B\to 0$ (in eq. \botto), i.e. when the extra
scalar $\v$ decouples from our model.

As a possible ansatz, we postulate that
$$
Q_2=\sqrt{{1\over 3} + \l(B,d)}
$$
with an arbitrary $\l$ satisfying
$$
\lim_{B\to 0} \l =0
\eqn\ccinque
$$
to grant the return to DDK solution.

For such a class of solutions, we can write down the corresponding value
of $Q$ and $\a$ as
$$
\eqalign{
&Q=Q_1= -{1\over \sqrt{3}} \left[ B\sqrt{1+3 \l} - \sqrt{(1+B^2)(3\l+25-d)}
\right]\cr
&\a=-{1\over 2\sqrt{3}}\left[
\sqrt{(1+B^2)(25-d+3\l)}-\sqrt{(1+B^2)(1-d+3\l)}\right]\cr}
$$

Then the string tension
$$
\Gamma_h = 2 (1-h) {Q\over 2\a} +2
$$
is given by
$$
\eqalign {
\Gamma = {1-h\over 12(1+B^2)}&\biggr\{ B \sqrt{1+B^2}
\Bigl[\sqrt{(1+3\l)(25-d+3\l)}+\sqrt{(1+3\l)(1-d+3\l)}\Bigr]+\cr
&-(1+B^2)\Bigl[25-d+3\l+\sqrt{(25-d+3\l)(1-d+3\l)}\Bigr]\biggr\}+2\cr}
$$

Among all these, we may further restrict ourselves by choosing
$$
\l={Bd\over 3}
$$
which satisfy the condition \ccinque.

Then one ends up with one-parameter family of solutions
$$
\eqalign{
&Q=Q_1=-{1\over \sqrt{3}}\left[B\sqrt{1+Bd}-\sqrt{(1+B^2)(25+(B-1)d)}\right]\cr
&\a=-{\sqrt{1+B^2}\over 2\sqrt{3}} \left[\sqrt{25+(B-1)d}-\sqrt{1+(B-1)d}
\right]\cr}
$$
and the string tension
$$
\Gamma_h={2 (1-h)\over \sqrt{1+B^2}}
{B\sqrt{1+Bd}-\sqrt{(1+B^2)(25+(B-1)d)}\ove
   r
\sqrt{25+(B-1)d}-\sqrt{1+(B-1)d}} +2
\eqn\nuovogamma
$$

We see that the string tension $\Gamma_h (B,d)$ is real for
$$
B\geq 1-{1\over d}
\eqn\realstring
$$
and with an arbitrary positive $d$.

We still have an unknown parameter $B$, whose physical meaning as well as
the possible values consistent with our model are unclear. We
would like to see that $B$ might exhibit the characteristic of an
"order parameter" with the $d$ dependent behaviour such as
$$
B=\bar B \theta (d-1)
\eqn\teta
$$

But it seems to be beyond the reach of the technic used so far in this work
to justify such a conjecture.

At this point, we may choose between two opposite points of view for going
further.

On one hand, one may try to supplement the current algebra type arguments of
foregoing
paragraphs with some sort of dynamical calculation and to see if one
obtains the "effective action" of the type of \botto,
along with the estimate (presumably perturbative) of unknown $B$.
One of the few works which seriously tries to set up the dynamical
calculational
scheme in 2D gravity is ref. [10], where perturbative scheme based on
$d=2+\epsilon$ expansion is constructed. It needs, however, further working
to set out the
consequences of such a scheme to answer questions such as these
mentioned above.

Another possibility is to take our "new" action \botto\ as starting point
and to see if such a model can give a
reasonable answer to some open physical problems of the DDK model.
One of the immediate questions one may ask to the model such as \botto\ is
the study of the
correlation functions for suitable physical operators (vertex
functions).
Here we have a set of well developed techniques which can be applied
just to the calculation of correlators of 2D gravity interacting with
matter (2D off-critical string) [11] [12].

In the rest of this paper, we propose to apply similar techniques to our model
and to see if there is some particular favourable range of values for the
parameter $B$, so that one may get non trivial (non zero) results for the
correlation functions of certain classes of vertices.
\vglue0.4cm
{\bf 4. Three and four points correlation functions
on the sphere}
\vglue0.6cm
We proceed to study tachyon scattering amplitudes of our model coupled
to a conformal matter represented by the following action
$$
S_{\picc M}= {1\over 8\pi} \int d^2 x \sqrt{\g} [- X^a \qgg X_a ]
\eqn\pa
$$
where $X^a$ are $d$ bosonic fields and the central charge for the matter sector
is $c_{\picc M}=d$.
Consistently with our interpretation of the area operator, matter vertex
$V_k=e^{ik_a X^a}$ get a gravitational dressing by the Liouville field alone
$$
T_k = e^{i k_a X^a + \b_1 \F^1}~~~~~~~~(k_a=k_{a'}~\forall a,a')
\eqn\pb
$$
where the parameter $\b_1=\b(k)$ is defined by the conformal invariance
($\Delta[T_k]=1$), that gives
$$
\b(B,k)=-{\sqrt{1+B^2}\over 2\sqrt{3}} \left [ \sqrt{25+d (B-1)}-
\sqrt{1+d (B-1) +12 d k^2}\right]
\eqn\pc
$$

We can express tachyon scattering amplitudes on the sphere
as follows [12] (the index $i$ in the
following indicates different vectors $k_i=\{k_{i1}, ..., k_{id}\}$, and must
not be confused with the component's index $1\leq a\leq d$ used above)
$$
\eqalign{
&A_n(k_1, ..., k_n):=<T_{k_1}, ..., T_{k_n}>=\cr
&={2\sqrt{\pi}\over \a_1} {\cal{A}}^{-1-s} \int D_{\g} \F' D_{\g} X'
e^{-S^{\picc 0}_{\picc grav}[\F'] -S^{\picc 0}_{\picc M}[X']}
\left [\int d x^2 \sqrt{\g}
e^{\a_1\F'^1(x)}
\right]^s T_{k_1} ... T_{k_n}\cr}
\eqn\pd
$$
where $\cal{A}$
is the area of the Riemann surface, $\F'$ and $X'$ represent the non
zero modes of the fields, $S^{\picc 0}$ are the quadratic part of actions,
$s$ is given by
$$
s= -{Q_1\over \a_1} - {1\over \a_1}\sum_{i=1}^n \b(k_i)
\eqn\chargea
$$
and charge neutrality of the matter sector requires
$$
\sum_{i=1}^n k_i =0
\eqn\chargeb
$$

In case of integer $s$ we can reduce \pd\ to the following multiple integral
$$
\eqalign{
&A_n={2\sqrt\pi\over \a_1} {\cal{A}}^{-1-s} \cr
=&\prod_{i=1}^n \prod_{j=1}^s
\int d^2 x_i d^2 t_j |x_i-t_j|^{-{2\over 1+B^2} \a_1 \b_i}
\prod_{i'<i} |x_{i'}-x_i|^{-{2\over 1+B^2}\b_{i'} \b_i +2 k_{i'} k_i}
\prod_{j'<j} |t_{j'}-t_j|^{-{2\over 1+B^2}\a_1^2}\cr}
\eqn\pe
$$

Eq. \pe\ represent the set of correlation functions of the $B$ dependent
infinite class of models defined by the actions in eq. \botto\ and \pa .
We now study this set of functions of the parameter $B$ to find a subset of
them
which show a good behaviour, i.e. which are convergent and not trivial
functions
of the momenta. In this way a minimum class of existing models with
physical meaning is achieved.

To reduce complexity we fix the value of $d$ to the most interesting case of
$d=4$ (the general case is explained in appendix C)
and we study \pe\ for $n=3$ and $n=4$, that is the three and four points
correlation functions.
We proceed in two steps. At first we analyse the integral in eq. \pe\ as
functions of $B$ to define the range of convergence in $B$-axis. Then we
explicit calculate the integral for fixed values of $B$ in the allowed
region.

For $n=3$, eq. \pe\ reduces to
$$
A_3(k_1, k_2, k_3) = {2\sqrt\pi\over \a_1}{\cal{A}}^{-1-s}
\prod_{i=1}^s \int d^2 t_i |t_i|^{2a} |1-t_i|^{2b} \prod_{i'<i}
|t_{i'}-t_i|^{4c}
\eqn\pf
$$
where
$$
\eqalign{
&a=-{1\over 1+B^2} \a_1\b(k_1)~~~~~~~b=-{1\over 1+B^2} \a_1\b(k_3)\cr
&c=-{1\over 2(1+B^2)} \a_1^2\cr}
$$
and the parameters are subjected to the charge conservation conditions
\chargea\ and \chargeb . In appendix A we
find the following convergence conditions
for the integrals \pf\
$$
\eqalign{
&Re~a, b>-1\cr
&-Min\left[ {1\over 2 (s-1)}; Re{a+1\over s-1};Re{b+1\over s-1}\right]<
Re~c<-Max\left[{1\over 2} Re {a+b+1\over s-1};Re{a+b+1\over s-1}\right]\cr}
\eqn\pg
$$

Evaluating these conditions for the parameters of our model
in function of $B$, we find the following result [14]: $A_3$ is well defined
for $s=1$ and $s=2$ respectively if $B\in [3/4;3/2+\e]$ and $B\in [3/2+\e;
8/5]$, while it does not exist any interval of $B$-axis for which $A_3$
is convergent for $s>2$.

For $n=4$, eq. \pe\ gives
$$
\eqalign{
&A_4(k_1, k_2, k_3, k_4) =\cr
&= {2\sqrt\pi\over \a_1} {\cal{A}}^{-1-s}
\int d^2 x |x|^{2a'} |1-x|^{2b'}
\prod_{i=1}^s \int d^2 t_i |t_i|^{2a} |1-t_i|^{2b} |x-t|^{2p}
\prod_{i'<i} |t_{i'}-t_i|^{4c}\cr}
\eqn\ph
$$
where
$$
\eqalign{
&a=-{1\over 1+B^2} \a_1\b(k_1)~~~~~~~b=-{1\over 1+B^2} \a_1\b(k_3)\cr
&p=-{1\over 1+B^2} \a_1\b(k_4)\cr
&a'=k_1 k_4-{1\over 1+B^2}\b(k_1)\b(k_4)~~~~b'=k_3 k_4-{1\over 1+B^2}
\b(k_3) \b(k_4)\cr
&c=-{1\over 2(1+B^2)} \a_1^2\cr}
$$

When $s=1$ in \ph , the convergence conditions are the following (appendix A)
$$
\eqalign{
&Re ~a,b,a',b'>-1\cr
&-Min\left[ {1\over 2}; Re {a+a'\over 2}+1; Re {b+b'\over 2}+1\right]
<Re{p\over 2}<\cr
&<-Max\left[ {1\over 2} Re(a+b+1); {1\over 2} Re (a'+b'+1);
{1\over 2} Re (a+b+a'+b'+1) +1\right]\cr}
\eqn\pl
$$

Now considering eq. \chargea\ and \chargeb\ and the condition $a'+b'<-1$,
necessary for \pl , by numerical evaluation we find that for $B\geq 3/2$ $A_4$
does not converge while for $B=3/4$ we can find a region in the $\{k_1,
k_2, k_3, k_4\}$ $4$-dimensional space where $A_4$ converges.

Above consideration allows to ignore the case $s=2$, because it would require
$B\geq3/2$ to ensure convergence of $A_3$. Then we can state that, even by
modifying $B$, we can not find meaningful correlation functions for $s>1$,
this fact, of course, suggest to fix $s$ to $1$.
This choice reduces the number of independent momenta, because of
conditions \chargea . Moreover
we note that a strong simplification arises in the
form of the parameters when $B=3/4$, since in this case they becomes linear
functions of (the moduli of) the momenta (see appendix C).
Then fixing the value of $B$ to
$B=3/4$, we define exactly one model for which we can now proceed to calculate
the three and four correlation functions and their relative regions of
convergence in the space of momenta.

As shown in appendix B, for the three points correlator we get
$$
\eqalign{
A_3(k_1, k_2, k_3)&= \cr
&={2\sqrt\pi\over \a_1} {\cal{A}}^{-2} \int d^2 t |t|^{2a}
|1-t|^{2b}\cr
&={2\sqrt\pi\over \a_1}  {\cal{A}}^{-2} \D (a+1) \D(b+1) \D(-a-b-1)\cr
&=-{4\over 5} \sqrt{2 \pi^3}  {\cal{A}}^{-2} \D(\sqrt{8} k_1-1)
\D(\sqrt{8} k_3 -1)
\D(\sqrt{8} k_2 +3)\cr}
$$
where the kinematic
has been fixed as $k_1, k_3>0,~k_2<0$, and conditions \chargea\ and \chargeb\
give
$$
\eqalign{
&k_2=-{1\over \sqrt{8}} \left(2+{\sqrt{6}\over 5}\right)\cr
&k_3=-k_2-k_1\cr}
$$

The region of convergence in the ${k_1, k_3}$ plane is given by
$$
\eqalign{
&k_1, k_3>{1\over\sqrt{8}}\cr
&k_1+k_3<{3\over \sqrt{8}}\cr}
$$
or in the $k_1$-axis by
$$
{1\over\sqrt 8}<k_1<{1\over\sqrt 8}\left( 1-{\sqrt 6\over 5}\right)
$$

For the four points correlator we must calculate
$$
A_4(k_1, k_2, k_3, k_4)={2\sqrt\pi\over \a_1}  {\cal{A}}^{-2} \int d^2 x
|x|^{2a
   '}
|1-x|^{2b'} \int d^2 t |t|^{2a} |1-t|^{2b} |x-t|^{2p}
\eqn\pp
$$
by choosing kinematic $k_1, k_3, k_4>0, ~k_2<0$, we get
$$
\eqalign{
&a=\sqrt{8} k_1-2~~~~~b=\sqrt{8} k_3 -2\cr
&p=\sqrt{8} k_4 -2\cr
&a'=-3 k_1 k_4 +\sqrt{8} (k_1+k_4)-2~~~~b'=-k_3 k_4 +\sqrt{8} (k_3+k_4)-2\cr}
$$

Imposing \chargea\ and \chargeb\ we reduce the free variable to $k_1$ and $k_4$
$$
\eqalign{
&k_2 = -{1\over \sqrt{8}} \left(3+{\sqrt{6}\over 5}\right)\cr
&k_3= -k_2-k_1-k_4\cr}
$$
so we can express the convergence region in the plane $\{k_1,k_4\}$ by
evaluating conditions \pl . The first set of inequalities gives
$$
k_i > {1\over \sqrt{8}}\sim 0.353553
$$
while in the second set of inequalities (the ones involving $p/2$), three
can be reduced in relations involving only $k_4$ that give the range
$$
0.353553<k_4<0.454869
$$
the other ones give the following system of conditions
$$
\eqalign{
&-3k_{1,3}k_4+\sqrt{8} (k_{1,3}+k_4) -1 >0\cr
&3 k_{1,3} k_4 -2\sqrt{8} (k_{1,3}+k_4)+4<0\cr}
$$
that can be solved numerically or geometrically showing that a region of
convergence exists around the value $k_{1,4}=0.4$.

We conclude by illustrating a calculation of \pp .
Consider the following function of the $5$ parameters $a, b, a', b', p$
$$
\eqalign{
I(a',b',a,b,p)&= \int d^2 x |x|^{2a'}
|1-x|^{2b'} \int d^2 t |t|^{2a} |1-t|^{2b} |x-t|^{2p}=\cr
&=\int_{-\infty}^{+\infty} du_1 dv_1 (u_1^2+v_1^2)^{a'}
[(1-u_1)^2 +v_1^2]^{b'}\cr
&~~\int_{-\infty}^{+\infty} du_2 dv_2 (u_2^2+v_2^2 )^a [(1-u_2)^2+v_2^2]^b
[(u_1-u_2)^2 +(v_1-v_2)^2]^p\cr}
$$

Making the two successive changes of variables [11]
$$
\eqalign{
&v_{1,2}\to i e^{-2i\e}\sim(1-2i\e) i v_{1,}~~~~~\e\to 0\cr
&x_{\pm} =u_1\pm u_2~~~~y_{\pm}=v_1\pm v_2\cr}
$$
we get
$$
\eqalign{
I(...)=-{1\over 4} \int_{-\infty}^{+\infty} &dy_+ dy_- [y_+-i\e (y_+-y_-)]^{a'}
[y_- +i\e(y_+ -y_-)]^{a'}\cr
& [y_+ -1 -i\e(y_+ -y_-)]^{b'}
[y_- + i\e(y_+ - y_-)]^{b'} G(a,b,c,x_+,x_-,y_+,y_-)}
\eqn\pp
$$
where
$$
\eqalign{
G(...)=&\int_{-\infty}^{+\infty} dx_+ dx_- [x_+ -i\e(x_+ -x_-)]^a
[x_- +i\e(x_+ -x_-)]^a\cr
& [x_+ -1 -i\e(x_+ -x_-)]^b [x_- -1+i\e(x_+ -x_-)]^b\cr
&[x_+ -y_+ -i\e(x_+ -x_- -y_+ +y_-)]^p [x_- -y_- +i\e(x_+ -x_- -y_+ -y_-)]^p
\cr}
$$

We can now proceed to calculate $I(a',b',a,b,c)$ using the fact that \pp\
factorizes in two parts depending on $+$ and $-$ variables when $\e\to 0$, and
the singularities of the integrand can be moved to the upper complex
half-plane by opportune choice of the regions of integration. More precisely,
$G(...)$ can be expressed as follows (see [11] where this technique
has been defined and exploited in a similar but different case)
$$
\eqalign{
G(...)
&=\Big[ (1-e^{-2\pi i a})\int_0^{y_+} dx_+ x_+^a (x_+ -1)^b \int_{-\infty}
^0 dx_- x_-^a (x_- -1)^b +\cr
&+ (1-e^{-2\pi ib}) \int_{y_+}^1 dx_+ x_+^a (1-x_+)^b
\int_1^{+\infty} dx_- x_-^a (x_- -1)^b \Big]\cr
&[x_+ -y_+ -i\e(x_+ -x_- -y_+ +y_-)]^p
[x_- -y_- +i\e(x_+ -x_- -y_+ +y_-)]^p\cr}
\eqn\ga
$$

In this formula the integration is of course intended also on the two
last factor common for each term, and the $\e$ dependence has been omitted
in the pieces depending only on $x$ variables as unnecessary for further
calculation. Note also that we have considered in this expression of $G$ only
the terms that will be non zero in successive $y$ integrations.

The method used to calculate $G(...)$ proceeds as follows. First we
divide the $x_+$ range of integration in the three region $(-\infty,0)$,
$(0,1)$ and $(1,+\infty)$. Then we evaluate the position of the singularity
of the $x_-$ integrand relatively to these regions. They are
$x_-=-i\e (x_+ -x_-)$, $x_-=1-i\e(x_+ -x_-)$  and $x_-=y_- -i\e(x_+ -y_+)$.
By evaluating
all the possible combinations, we see immediately that the only non zero terms
are the one in eq. \ga\ plus two other terms
that anyway will disappear because of further integration in $y$.
In fact, for example, if $x_+ \in (-\infty ,0)$ and $x_+ <y_+$
all the singularities
are in the upper half-complex plane, so we can close the path of integration
without crossing the branching line of the integrand and the result is zero.
{}From the interval $(0,1)$ we get the two terms of eq. \ga , and finally from
the first and the last interval we
get two terms that will be eliminated by further
integration. In the third step, in fact, to manage the $y$ singularities,
we part the $y_+$ range of
integration in the three intervals as above. By reconsidering all the possible
combinations and by taking in account that now the singularity depending from
the sign of $x_+ -y_+$ change, we readily obtain that the only surviving terms
are the ones relative to the $y_+ \in (0,1)$ range, so we get
$$
\eqalign{
&I(a',b',a,b)=\cr
&=-{1\over 4}\Big[ (1-e^{-2\pi ib'})(1-e^{-2\pi ia}) \int_0^1 dy_+
y_+^{a'} (y_+ -1)^{b'} \int_0^{x_+} dx_+ x_+^a (x_+ -1)^b (x_+ -y_+)^p\cr
&\int_1^{+\infty}dy_- y_-^{a'} (y_- -1)^{b'} \int_{-\infty}^0 dx_- x_-^a
(x_- -1)^b (x_- -y_-)^p +\cr
&+ (1-e^{-2\pi ib})(1-e^{-2\pi ia'}) \int_0^1 dy_+
y_+^{a'} (y_+ -1)^{b'} \int_{x_+}^1 dx_+ x_+^a (x_+ -1)^b (x_+ -y_+)^p\cr
&\int_{-\infty}^0 dy_- y_-^{a'} (y_- -1)^{b'} \int_1^{+\infty} dx_- x_-^a
(x_- -1)^b (x_- -y_-)^p \Big]\cr}
\eqn\gb
$$

By simple considerations we can transform the second term obtaining
$$
I(a',b',a,b,c)=I_1(a',b',a,b,c)+I_1(a,b,a',b',c)
$$

Now a series of change of variables and the extraction of a phase factor,
together with the possibility of reformulating the $(-\infty, 0)$ integral as
a linear combination of integrals on $(1, +\infty)$ and $(1,y)$ (see [11] for
details), lead to the following result
$$
\eqalign{
I_1 (a',b',a,b,c)&=s(b')s(b+p) A(a',b',a,b,c) A'(a',b',a,b,c)+\cr
&+s(b) s(b')
A(a',b',a,b,c)A'(a,b,a',b',c)\cr}
$$
where
$$
\eqalign{
&A'(a',b',a,b,c):=A(-a'-b'-p-2,b',-a-b-p-2,b,c)\cr
&A(a',b',a,b,c):=\int_0^1 dt~ t^{a'} (1-t)^{b'} \int_0^t ds~ s^a (1-s)^b
(t-s)^p\cr
&s(x):=sin(\pi x)\cr}
$$
and so
$$
\eqalign{
I(a',b',a,b,c)&=s(b')s(b+p) A(a',b',a,b,c) A'(a',b',a,b,c)+\cr
&+s(b)s(b'+p)
A(a,b,a',b',c)A'(a,b,a',b',c)+\cr
&+s(b) s(b')[A(a',b',a,b,c)A'(a,b,a',b',c)+
A(a,b,a',b',c)A'(a',b',a,b,c)]\cr}
$$

The last step consists in calculating $A(a',b',a,b,c)$, we can proceed as
follows
$$
\eqalign{
&A(a',b',a,b,c)=\cr
&={\G(a+1)\G(b+1)\over\G(a+p+2)} \int_0^1 dt t^{a'+a+p+1}
(1-t)^{b'} F(a+1,-b;a+p+2;y)=\cr
&={\G(a+1)\G(b+1)\over\G(a+p+2)}
{\G(a+a'+p+2)\G(b'+1)\over \G(a+a'+b'+p+3)}\cr
&~~~_3 F_2 (a+1,-b,a+a'+p+2;a+p+2,a+a'+b'+p+3;1)=\cr
&={\G(p+1) \G(b'+1)\over \G(-b)} \sum_n {\G(a+1+n) \G(-b+n) \G(a+a'+p+2+n)
\over \G(a+p+2+n) \G(a+a'+b'+p+3+n)}\cr}
$$

The last remark on this result is that the series converges when the conditions
\pl\ are satisfied, in fact we can evaluate the behaviour of the term $\s_n$
of the series as follows [15]
$$
\s_n={\G(p+1) \G(b'+1)\over\G(-b)} n^{-b-b'-p-4} [1+O(n^{-1})]
$$
and so the series converges absolutely if $Re~p>-Re(b+b')-3$,
that is weaker than \pl .

\vglue0.4cm
{\bf 5. General model}
\vglue0.6cm
The considerations of previous section is sufficient to convince us the
existence of a certain minimal subset of models, of our theory,
parameterized by $B$, which are "physically non trivial",
i. e. with non-zero three and four points
correlation functions. Moreover, the behaviour
of n-points function is such that, with the increasing value of n (actually
we have checked only from n=3 to n=4), the allowed value for $B$ is "driven"
to its minimal possible value for ensuring the reality of the string
susceptibility, i. e. $B=3/4$ for $d=4$.
We can also consider another strong argument that
lead to the same choice of $B=3/4$ for the representative model. This argument
starts from a consideration on the form of the string susceptibility $\G$
for generic $d$ (eq. \nuovogamma )
$$
\Gamma_h={2 (1-h)\over \sqrt{1+B^2}}
{B\sqrt{1+Bd}-\sqrt{(1+B^2)(25+(B-1)d}\over
\sqrt{25+(B-1)d}-\sqrt{1+(B-1)d}} +2
$$
where $B$ is in general a function of $d$ for which the only condition is
$$
\lim_{d\to 1} B(d) = 0
\eqn\lim
$$
(this is essentially a different way of formulating conditions \ccinque\
and \teta\ of
previous session).

This condition alone obviously does not allow to determinate the analytic form
of $B$, but if we remember that the condition for reality of $\G$ is
\realstring\
$$
B\geq 1-{1\over d}
$$
and keep in account the results of previous section, that can be read as
$$
1-{1\over d}\leq B\leq 1-{1\over d}+\e ~~~~~~d=4;~~~\e\to 0
$$
then we can guess the following analytic form for $B=B(d)$
$$
B(d)=\left( 1-{1\over d}\right) \t (d-1)
$$
which of course satisfies \lim\ (this choice of $B$ correspond in fact in
setting in \teta\ $\bar B=1-1/d$).

In such a way we obtain the following simple form for string susceptibility
$\G=\G(d)$, say in genus zero ($h=0$)
$$
\G(d)=\left\{
\eqalign{&{1\over \sqrt{6}} \sqrt{{d\over 2d^2-2d+1}} (d-1),~~d\geq 1\cr
&{1\over 12} \left[ d-1-\sqrt{(25-d)(1-d)}\right],~~d\leq 1\cr}\right.
$$

If we study this function of $d$, we find the graph in figure 1.
In this way, we obtain
a continuum evaluation of $\G$ for all values of $d$. In particular the value
of $\G$ for $d=3$, exceeds the theoretical upper bound ($\G={1\over 2}$),
predicted by J. Fr$\ddot o$hlich [16] and coworkers (who consider as dominant
in this range the surfaces degenerated in branched polymer), but in good
agreement with the recent numerical results of M. Caselle et al.
[17] (who conjecture
the possibility of avoiding the problem of crumbled surfaces by reconsidering
the whole sum over all genera).

\vglue0.4cm
{\bf 6. BRST quantization}
\vglue0.6cm
The possibility of constructing the consistent quantum theory of 2D (induced)
gravity depends on its BRST structure, corresponding to the combined symmetries
under diffeomorphism and Weyl transformations.

On the series of works [18], [19] and [20], R. Stora and coworkers have given
the complete treatment of this aspect of the on and off critical string theory.

However, there seems to be still some confusions in the existing literatures
about the correct formulation, in particular on the relationship between
diffeomorphism and Weyl symmetries and consequent introduction of Liouville
field "$\f_{\picc L}$". On ref. [18], [19] and [20],
some of the relevant results
are only indicated rather than fully worked out.

To state carefully our point of view on the subject (which is fully in
accordance with that of above quoted references), we summarize the relevant
results of above works and apply it to the case of particular type of
(Weyl) gauge fixing which interests us in the present work.

The BRST symmetry concerned comes from

{\it 1)} diffeomorphism (reparameterization invariance of the 2D surface)

$$
\eqalign{
&z\to f(z,\z)\cr
&\z\to\bar f(z,\z)\cr}
$$

{\it 2)} Weyl symmetry

$$
g_{ab}\to e^{\s}g_{ab}  ~~~~~(\s~ local~function)
$$

Let us take, following ref. [21], as the $0^{\picc th}$ order (classical)
action, the $d$-bosonic matter interacting with 2D gravitational field
represented by the metric $g_{ab}$
$$
S_0 = \sum_{\mu=0}^d \int d^2 x \sqrt{g} g^{ab} \nabla_a X_{\mu}
\nabla_b X_{\mu}
$$

It is convenient to parameterize the 2D metric $g_{ab}$ in terms of Beltrami
differentials [21].
The invariant line element is given by
$$
ds^2=e^{\phi}|dz+\mu d\z|^2
$$
with
$$
\eqalign{
&g_{\z\z}=\mu e^{\phi},~~g_{zz}=\bar\mu e^{\phi}\cr
&g_{z\z}=g_{\z z}={1+\mu\bar\mu\over 2}e^{\phi}\cr}
$$
and
$$
\sqrt{g}=e^{\phi} {1-\mu\bar\mu\over 2}~~~~(\mu\bar\mu<1)
$$

In term of these parameters, the action $S_0$ takes the form
$$
S_0=\sum_{\mu=1}^d \int_{\Sigma} {dz\land d\z\over 2i}
{(\d_z -\bar\mu\d_{\z}) X_{\mu} (\d_{\z} -\mu\d_z) X_{\mu} \over
(1-\mu\bar\mu)}
$$

Clearly $S_0$ is invariant under the symmetries {\it 1)} and {\it 2)}
(Note that $S_0$ does
not contains the Weyl factor $\phi$. One can say that $S_0$ depends only on
the complex structure of $\Sigma$ [23]).

The BRST transformations corresponding to the symmetry (diffeomorphism $\times$
Weyl) are expressed in terms of diffeomorphism ghosts $\xi,\bar\xi$, and
Weyl ghost $\psi$. In view of partial holomorphic factorization
of diffeomorphism part of the symmetry, it is convenient also to introduce
the auxiliary ghosts fields [20]
$$
\eqalign{
&C=\xi+\mu\bar\xi\cr
&\bar C=\bar\xi+\bar\mu\xi\cr}
$$
then the BRST transformation of relevant variables are
$$
\eqalign{
&s X_{\mu}=(\xi\cdot \d) X_{\mu}\cr
&s\mu=\bar\d C-\mu \d C+\d\mu C\cr
&s \phi=\psi+(\xi\cdot\d)\phi+\d\xi\mu\d\bar\xi+\bar\mu\bar\d\xi\cr
&\xi=(\xi\cdot\d)\xi\cr
&s C=C\d C\cr
&s\psi=(\xi\cdot\d)\psi\cr}
$$
where $(\xi\cdot\d)$ stands for $\xi\d+\bar\xi\bar\d$.

For the gauge fixing condition, we choose conventional conformal gauge
$$
\eqalign{
&\mu=\mu_0\cr
&\bar\mu=\bar\mu_0\cr}
$$
with respect to diffeomorphism but leave the condition for Weyl gauge open
for the moment
$$
F(\mu,\bar\mu,\phi)=0
$$

The gauge fixing part of the action is then
$$
S_{\picc g.f.}= \int_{\Sigma}{dz\land d\z\over 2i}
s[B(\mu-\mu_0)+\bar B (\bar\mu-\bar\mu_0)+\bar\psi F(\mu,\bar\mu,\phi)]
$$

The auxiliary fields $B,\bar B$ and $\bar\psi$ transform as
$$
\eqalign{
&s b =B\cr
&s \bar b=\bar B\cr
&s\bar\psi=D\cr
&s B=s\bar B=s D=0\cr}
$$
and, naturally, we have $s^2=0$. (Diffeomorphism anomaly at one loop [19]).
The first problem for the consistent quantization of off critical string
is that the Slavnov-Taylor identity for the effective action (or
"vertex functional")
$$
\Gamma=S_0+S_{\picc g.f.}+S_{\picc sources}+\hbar \Gamma_1 +....
$$
becomes obstructed at one loop level. Indeed, the term
$\Gamma_1$ contains the one loop integral over $X_{\mu}$'s as well as ghosts
$b$
and $c$. As it is well know, such a contribution takes the form
$$
\Gamma_1'=\gamma'(W(\mu)+\bar W(\bar\mu))
$$
with $\gamma'\propto d-26$.
While the actual form [1] [26] of one loop integral $(W(\mu)+\bar W(\bar\mu))$
does not matter here, it is important to note that $\Gamma_1'$ does not satisfy
the Slavnov identity at one loop level
$$
\hat\beta\Gamma_1'\equiv\int[\e(\bar\d-\mu\d-2\d\mu)+\bar\e(\d-\bar\mu\bar\d-2
\bar\d\bar\mu)]\times\G_1'\equiv 0
$$

In terms of the holomorphic component $W(\mu)$, this would amount to
$$
(\bar\d-\mu\d-2\d\mu) W\equiv 0
$$

Instead, we have an anomaly in RHS
$$
(\bar\d-\mu\d-2\d\mu) W={1\over 12\pi }\d_z^3 \mu
\eqn\zp
$$

To maintain the invariance under the diffeomorphism and thus to guarantee
the consistent quantization of the model, it is necessary to introduce
the one loop counter term $\L (\mu,\bar\mu,\phi')$ (Wess-Zumino term) to
cancel the anomaly \zp\
$$
\hat{\beta} (\G_1'+\L)=0
$$

We introduce the Wess-Zumino field $\phi'$ with the BRST transformation
given by
$$
s \phi'=(\xi\cdot\d)\phi'+\d\cdot\xi+\mu\d\bar\xi+\bar\mu\bar\d\xi
$$

The explicit form of local counter term $\L$ is again well known [19] [27]
$$
\eqalign{
\L(\mu,\bar\mu,\phi')=&\cr
= -{\gamma'\over 2}\int {dz\land d\z\over 2i}
{1\over 1-\mu\bar\mu} [(\d-\bar\mu\bar\d)\phi'(\bar\d-\mu\d)\phi'
&-2(\bar\d\bar\mu(\bar\d-\mu\d)+
\d\mu(\d-\bar\mu\bar\d))\phi'+F(\mu,\bar\mu)]\cr}
$$
where $F(\mu,\bar\mu)$ is the local function of the $\mu$ and $\bar\mu$ only.

Making use of the explicit expression of the scalar curvature in terms of
Beltrami's differential, one can also write $\L$ as
$$
\L(\mu,\bar\mu,\phi')= -{\gamma'\over 2}\int {dz\land d\z\over 2i}
{1\over 1-\mu\bar\mu} [(\d-\bar\mu\bar\d)\phi'(\bar\d-\mu\d)\phi'-
\sqrt{\tilde g}\tilde R\phi'+F(\mu,\bar\mu)]
$$

The new field $\phi'$ can be related to the Liouville field $\phi_{\picc L}$
\footnote{*}{The introduction of "new" Liouville field $\f_{\picc L}$
(or $\f$') is necessary except for $d=26$. See K. Fujikawa [24].}
such as introduced in ref. [19] with transformation
$$
s\phi_{\picc L}=\psi+(\xi\cdot\d)\phi_{\picc L}
$$

One has
$$
\phi'=\phi_{\picc L}-\phi
\eqn\zv
$$
($\phi$ is the Weyl factor)

The presence of $\phi'$ with an action $\L(\mu,\bar\mu,\phi')$, implies
the further renormalization of overall coefficient $\gamma'$. Because of the
one loop contribution from $\phi'$, one has
$$
\G_1'={d-26\over 24\pi} (W(\mu)+\bar W(\bar\mu))\to{(d+1)-26\over 24\pi}
 (W(\mu)+\bar W(\bar\mu))
$$
i.e. $\phi'$ behaves like $X_{d+1}$.

Thus the necessary counter term is now [19] [25]
$$
\L'(\mu,\bar\mu,\phi')= {d-25\over 24\pi}\int {dz\land d\z\over 2i}
{1\over 1-\mu\bar\mu} [(\d-\bar\mu\bar\d)\phi'(\bar\d-\mu\d)\phi'-
\sqrt{\tilde g}\tilde R\phi'+F(\mu,\bar\mu)]
$$

Further the presence of linear term $\sqrt{\tilde g} \tilde R\phi'$
necessitates the term which acts to cut off the large positive fluctuations
of $\phi'$. Thus one introduces the cosmological term
$$
\mu_0 \int {dz\land d\z\over 2i}\sqrt{\tilde g}\tilde R e^{-\phi'}
$$
which is s-invariant.

Ref. [18] [19] [20] choose the simplest Weyl gauge fixing
$$
\phi=\phi_0 ~~~(i.e.~~F(\mu,\bar\mu,\phi)=\phi-\phi_0)
$$

In this case, one can integrate out trivially the Weyl degrees of freedom
such as $\phi,\psi,\bar\psi$ and $D$, through algebraic conditions, leaving
essentially only the Liouville field $\phi'\sim\phi_{\picc L}$ as dynamical.
If, at this stage, one calculates the complete effective action [29] of the
model, one would end up in David-Distler-Kawai (DDK) action [2].

We are interested to the case (Weyl gauge fixing through non algebraic
equation)
$$
F(\mu,\bar\mu,\phi)=R(g)-R_0
$$
The corresponding gauge fixing action is
$$
S_{\picc g.f.}^{\picc Weyl}=\int {dz\land d\z\over 2i} \sqrt{g}[D(R(g)-R_0)
-\bar\psi{\dd\over\dd\s}R(e^{\s}g)|_{\s=0}\psi]
\eqn\zzc
$$

In this gauge, the Weyl factor $\phi$ enters explicitly into the classical
(tree-level) action through
$$
g=e^\f \tilde g
$$

As we have seen in the first section, one can write \zzc\ as
$$
S_{\picc g.f.}^{\picc Weyl}=\int {dz\land d\z\over 2i} \sqrt{g}[-D(
\quadgg\phi-R(\tilde g))
-\bar\psi(\quadgg-\tilde\quadg\phi+R(\tilde g))\psi]
$$

In terms of Beltrami differentials
$$
\eqalign{
S_{\picc g.f.}^{\picc Weyl}=&\int {dz\land d\z\over 2i}
\Big\{ {(\d-\bar\mu\bar\d)D(\bar\d-\mu\d)\phi\over 1-\mu\bar\mu}
+DR(\tilde g)+{(\d-\bar\mu\bar\d)\bar\psi(\bar\d-\mu\d)\psi\over 1-
\mu\bar\mu}\cr
&-\bar\psi\psi\left(R(\tilde g)-
[\d-\bar\d(\bar\mu]{1\over 1-\mu\bar\mu}[\bar\d-\mu\d]\phi\right)\Big\}\cr }
\eqn\zzf
$$

It is perhaps instructive, however, to first consider
the simpler gauge fixing mentioned in [22].
This amounts to consider
$$
\tilde\quadg\phi=0
$$
instead of $R(g)=R_0$.

The gauge fixing action is
$$
S_{\picc g.f.}^{\picc Weyl}=\int {dz\land d\z\over 2i} s(\bar\psi\tilde\quadg
\phi\sqrt{g})=\int {dz\land d\z\over 2i} [D\tilde\quadg\phi-\bar\psi\tilde
\quadg\psi]\sqrt{g}
$$

For $\mu_0=\bar\mu_0=0$ (diffeomorphism gauge fixing)
$$
S_{\picc g.f.}^{\picc Weyl}\sim\int {dz\land d\z\over 2i} (D\d\bar\d\phi
-\bar\psi\d\bar\d\psi)
$$

Now it is easy to see, as in ref. [22],
that the new propagating fields $D, \phi,
\bar\psi$ and $\psi$ (free) contribute to the central charges $+1,+1,-1$
and $-1$, respectively.

Thus, in this case, the gauge fixing in "diffeomorphism" discussed previously
is unaffected. In particular, one loop integrals corresponding to $D, \phi,
\bar\psi$ and $\psi$, potentially again diffeomorphism anomalous, cancel
out between themselves. In the end, one again ends up with the effective action
of DDK type.

Now the gauge fixing through curvature, \zzc -\zzf , introduces a set of
interactions between these modes $D, \phi, \bar\psi$ and $\psi$. Thus,
at the level of fundamental Lagrangian and corresponding Feymann
graph, it would be quite complicated to discuss the gauge invariance.
However, at the level of effective action [29], one can make appeal to the
basic
symmetry principle directly, and it is easy to write down the correct
form of the effective action involving new propagating modes $\psi$ and
$\bar\psi$.

Again to maintain the symmetry, replacing the Weyl factor $\phi$ with Liouville
field $\phi'$ (more precisely, according to \zv , $\phi'\sim\phi_{\picc L}-
\phi$) one obtains the additional terms of the form
$$
\int {dz\land d\z\over 2i} \sqrt{g}[-D(
\quadgg\phi'-R(\tilde g))
-\bar\psi(\quadgg-\tilde\quadg\phi'+R(\tilde g))\psi]
\eqn\zzl
$$
coming from the $S_{\picc g.f.}$ in original Lagrangian.

The first term of \zzl , however, corresponds to the gauge condition in
functional integral
$$
\dd (R(\mu,\bar\mu,\phi)-R_0)
$$

If we may anticipate the fact that the theory is truly gauge invariant under
diffeomor\-phism\-$\times$Weyl, then the physical result should not be affected
by the change
$$
R_0\to R_0+f(z,\z)
$$
where $f(z,\z)$ is an arbitrary real, fast falling function. Thus one may
replace the factor $\dd (R-R_0)$ by
$$
\int Df e^{-F(f)}\dd (R-R_0-f)
$$

Choosing, just as in the case of Yang-Mill theory,
$$
F(f)={1\over 2\a}\int \sqrt{g} f^2  {dz\land d\z\over 2i}
$$
the partition function will acquire the factor
$$
e^{-{1\over 2\a}\int (R-R_0)^2 \sqrt(g)}
$$
instead of the $\dd$ function. Clearly from the perturbation calculus
point of view this is much simpler than BRST form \zzc .

Thus, instead of \zzl , the relevant term in our effective action should be
$$
\int {dz\land d\z\over 2i}\sqrt{\hat g}[-{1\over 2\a}(R(\phi',\mu_0,\bar\mu_0)
-R_0)^2-\bar\psi(\quadgg-\quadgg\phi'+R(\hat g))\psi]
\eqn\zzp
$$

One must also suitably renormalize the coefficients in "DDK" Liouville term.
Note that the resultant action is of interacting type with "dynamical" degrees
of freedom $\phi', \psi$ and $\bar\psi$. According to the general rule
of effective action [29], one may compute the higher order term applying
formal Feymann rules according to \zzp , as if it was a fundamental lagrangian.
In section 2, we have shown that
the Liouville $+$ "Weyl determinant" part can be reduced to the gaussian
(quadratic) form. Thus the only interactions comes from curvature $(R-R_0)^2$
term.

Note that this term can be written as
$$
-{1\over 2\a}\int (R-R_0)^2\sqrt{g}=-{1\over 2\a}
\int R^2 \sqrt{g} +{32\pi^2 (1-h)^2\over \a {\cal{A}}}
$$
due to the relationship
$$
R_0(const)={8\pi (1-h)\over {\cal{A}}}
$$

Recently Kawai and Nakayama [28] has studied the model with effective action
of the form
$$
-[(DDK)+{1\over 2\a}\int R^2\sqrt{g}]
$$

In particular, they have shown that such a model has the small area scaling
behaviour
$$
\sim exp\left(-{32\pi^2 (1-h)^2\over \a{\cal{A}}}\right)
$$

In the present model, however, such a factor is exactly cancelled, leaving
presumably an ${\cal{A}}$-indipendent behaviour for small ${\cal{A}}$.
Anyway, for large $\cal{A}$, we must take simultaneously into account the
influence of Weyl jacobian and this should reproduce the scaling behaviour
discussed in section 3 and 4. Kawai and Nakayama shows that, in their model,
$R^2$ term does not influence large $\cal{A}$ scaling behaviour, thus
reproducing KZP formula for their model.
The fuller BRST study of our model, including physical spectrum, is in
progress.
\vglue 0.6cm
{\bf 7. Conclusions and discussions}
\vglue0.4cm
At first sight, it looks puzzling that we obtain new results by the
mere consideration of Weyl gauge fixing. All the more since, as it has
been argued in section 6, the "traditional" DDK result comes from a
consistent gauge fixing and moreover it gives the good result for $d<1$.

We interpret this situation as due to the incomplete aspect of our continuum
approach, i.e. the lack of a consistent computational scheme for strong
coupling regime which forces us to use a semiclassical approach "corrected"
with consistency arguments based on symmetries.
The fact that a particular choice
of gauges might result more "suitable" than others to formulate
a continuum string theory for $d>1$, is then not so strange.

In fact the consequent appearance of the $R^2$ term in our effective action
discussed in the previous section,
seems to correlate our approach with the arguments in ref. [28] for the
possible
not dominance of the branched polymer configurations at $d>1$.
This fact is supported by the scaling behaviour obtained, which is in
agreement with the picture of smooth
random surfaces embedded in a generic $d$-dimensional flat space-time.
In other works, contrary to some of the previous approaches, it looks as if the
   effect of
the class of surfaces with spiky configurations are suppressed
\footnote{*}{
Recently, M. Caselle et al.
has performed the simulation on the scaling
behaviour of inter-phase surface in 3-$d$ Ising model [17].
Their result supports
our formulation of string tension. We thank M. Caselle for the discussion
on this point.
}.

It should be noted that the addition of the
new "Weyl" scalar $\v$ to the original
DDK approach actually shifts the role of critical parameter from the
dimension $d$
to a new "dynamical" parameter $B$, which is the measure of coupling
between Liouville field $\phi$ and Weyl field $\v$. Indeed, the original
"phase transition" occurring at $d=1$, is now related with the varying of the
parameter $B$. Thus the numerical analysis given in section 4 and 5 should be
understood as an indication of the fact that our model is lying
in a neighbourhood of its critical line
$$
B=B_{\picc c}=1-{1\over d}
$$
for any arbitrary value of $d>1$.

Following up this idea, our next attempt should be the calculation of the
relevant critical exponents and correlations. The work is in progress.
\vglue 0.6cm
{\bf Acknowledgment}
\vglue0.4cm
M. Martellini is grateful to the hospitality rented to him at Dipartimento
di Fisica, Universit\`a di Roma "La Sapienza". K. Yoshida acknowledges the many
illuminating discussions with his japanese colleagues, in particular T. Eguchi,
K. Fujikawa, K. Higashiyama, H. Kawai and others. It is a pleasure
to thank prof. K. Igi and prof. H. Sugawara for hospitality. We also
acknowledge
the partial support from the Italian Ministry of Research and University, MPI
40\%.
\vglue 0.6cm
{\bf  Appendix A \hfil}
\vglue 0.4cm
We consider the integral:
$$
I[a,b,c,s]:= \prod_{i=i}^s \int_{C_{i}} d^2 z_i |z_i|^{2a} |1-z_i|^{2b}
\prod_{j<i} |z_j-z_i|^{4c}
\eqn\uno
$$
that can be written in the following alternative ways:

cartesian coordinates
$$
I[a,b,c,s]:= \prod_{i=i}^s \int_{-\infty}^{+\infty} dx_i dy_i
(x_i^2+y_i^2)^a [(1-x_i)^2+y_i^2]^b \prod_{j<i} [(x_j-x_i)^2+(y_j-y_i)^2]^{2c}
$$

polar coordinates
$$
I[a,b,c,s]:= \prod_{i=i}^s \int_0^{2\pi} d\t_i \int_0^{+\infty} dr_i r_i^{2a+1}
(1+r^2_i+2r_i cos\t_i)^b \prod_{j<i} [r_j^2+r_i^2+2r_j r_i cos(\t_j-\t_i)]^{2c}
$$

Convergence of integral \uno\ depends on the behaviour of the integrand in
function of the parameters $a, b, c, s$ in the singular points, which can be
identified in $z_i=0\ne z_k~ \forall k\ne i;~ z_i=1\ne z_k~ \forall k\ne i;~
z_i=+\infty\ne z_k~ \forall k\ne i.~$ Then also the cases when one or more
different variables assume the same value, as for example:
$z_i=z_k\ne 0,1,\infty~ \forall k;~ z_i=0~ \forall i;~ z_i=1~ \forall i;~
z_i=\infty~ \forall i$.

We start by simplicity to consider the case of $s=1$ and $s=2$.

For $s=1$ integral \uno\ becomes
$$
\eqalign{
I[a,b,c,1]&=  \int_{C} d^2 z |z|^{2a} |1-z|^{2b}\cr
&= \int_{-\infty}^{+\infty} dx~ dy(x^2+y^2)^a [(1-x)^2+y^2]^b\cr}
$$
and can be calculated (see appendix B) under the conditions:
$$
Re~a>-1,~Re~b>-1,~Re(a+b)<-1
$$

We can obtain the same conditions by considering the behaviour in
$z=0$ and $z=1$
$$
z=0 ~~~I\sim z^{2a+2}
$$
where the second term ($2$) in the exponent comes from the fact that we
integrate over two variables. From this behaviour we get the condition
$$
Re~a>-1
$$

Then by using the symmetry on the variable change $z\to 1-z$,
we can evaluate the behaviour in $z=1$ getting the condition for $b$,
$$
Re~b>-1
$$

In $z=\infty$ we have
$$
z=\infty, ~~~I\sim z^{2a+2b+2}
$$
from which we get the last condition $Re(a+b)<-1$.

For $s=2$ integral \uno\ becomes
$$
\eqalign{
I[a,b,c,2]&=  \int_{C} d^2 z_1 d^2 z_2 |z_1|^{2a} |z_2|^{2a}
|1-z_1|^{2b}
|1-z_2|^{2b} |z_1-z_2|^{4c}\cr
&= \int_{-\infty}^{+\infty} dx_1 dy_1 dx_1 dy_1 (x_1^2+y_1^2)^a
(x_2^2+y_2^2)^a
[(1-x_1)^2+y_1^2]^b  [(1-x_2)^2+y_2^2]^b\cr
&~~~~~~ [(x_1-x_2)^2+(y_1-y_2)^2]^{2c}\cr}
$$

Let proceed to evaluate the behaviour in the singular points. If just one of
the variable is zero, integral behaviour is the same as in the case above
from which the condition
$Re~a>-1$, in the same way for $z_i =1$ we can deduce the condition
$Re~b>-1$.
If just one of the variable goes to infinity we have
$$
z_1 =\infty,~z_2=k,~~~I\sim (z_1 z_2)^{2a+2b+4c+2}
$$
(from now on $k$ represents a finite constant different from zero and one)
from which the condition
$$
Re~c<-{1\over 2}Re(a+b+1)
\eqn\cc
$$

Let now evaluate what happens when both the variables have singular values.
If both the variables are zero, integral behaviour is the following
$$
z_1=0,~z_2=0~~~I\sim (z_1 z_2)^{2a+2a+4c+2+2}
$$
from which we deduce the condition
$$
-Re(a+1)<Re~c
$$
and as above, for $z_1=z_2=1$ we get the condition
$$
-Re(b+1)<Re~c
$$

If both the variables go to infinity, we have
$$
z_1=z_2=\infty,~~~I\sim (z_1 z_2)^{2a+2a+2b+2b+4c+4}
$$
from which the condition
$$
Re~c<-Re(a+b+1).
$$

If, finally, both the variables have
the same not singular value,
we get
$$
z_1=z_2=k,~~~I\sim (z_1- z_2)^{4c+2}
$$
in this case the term $2$ comes from the fact that singularities
occurs from values not of the variables but of their differences, that is
there are two, not four degrees of freedom. From this we get the following
condition
$$
Re~c>-{1\over 2}
$$

We can resume all the conditions as follows
$$
-Min\left[{1\over 2},Re(a+1),Re(b+1)\right]<Re~c<-Max \left[
{1\over 2}Re(a+b+1), Re(a+b+1)\right].
$$

We can then extend the above reasoning to the general case of arbitrary $s$.
When just one variable is $0$ or $1$ we get as before $Re~a,b>-1$.
When one variable is $\infty$, we get
$$
2a+2b+(s-1) 4c +2<0~~~\to~~~Re~c<-{1\over 2 (s-1)} Re (a+b+1)
\eqn\tr
$$

Then we must consider when $n\leq s$ variables are equal to $0$ or $1$,we get
$$
2na+{1\over 2}n(n-1)4c+n>0~~~\to~~~Re~c>-Re{a+1\over n-1}~~\forall n\leq s
$$
but since $a+1>0$, then the stronger condition is the one with $n=s$.
Similarly if $n\leq s$ variables are equal to $\infty$, we have
$$
n(2a+2b)+{1\over 2}n(n-1)4c+n<0~~~\to~~~Re~c<-{1\over n-1} Re(a+b+1) ~~\forall
n\leq s
$$
and we can see that we can reduce all this set of inequalities
together with the \tr\ to the following one
$$
Re~c <-Max\left[ {1\over 2(s-1)}(a+b+1);{1\over s-1} (a+b+1)\right]
$$

Finally for the cases when $n\leq s$ variables are equal but not
singular we get
$$
{1\over 2}n(n-1)4c+n>0~~~\to~~~Re~c>-{1\over 2(n-1)}~~\forall n\leq s
$$
that can be reduced to
$$
Re~c>-{1\over 2(s-1)}
$$

Then we can resume all these condition obtaining the ones stated in eq. \pg .

Note that the above technic can be applied to the case of integral in eq. \ph ,
when $s=1$, that is
$$
I(a',b',a,b,p)=\int d^2 x |x|^{2a'} |1-x|^{2b'}
 \int d^2 t |t|^{2a} |1-t|^{2b} |x-t|^{2p}
$$
and gives the conditions stated in eq. \pl .
\vglue 0.6cm
{\bf Appendix B \hfil}
\vglue 0.4cm
We can calculate directly the integral in the case $s=1$,
$$
\eqalign{
I[a,b,c,1]&=  \int_{C} d^2 z |z|^{2a} |1-z|^{2b}\cr
&= \int_0^{2\pi} d \t\int_0^{+\infty} r^{2a+1} (1+r^2+2rcos\t)^b\cr}
$$

We can rewrite the integral by using the following propriety of $\G$ function
$$
\G(x)=s^x\int_0^{+\infty} dt~e^{-st}t^{x-1}~~~~Re~s>0,~~Re~x>0
$$
under the conditions $Re~b<0$ we get
$$
\eqalign{
I(a,b)&={1\over  \G(-b)} \int_0^{+\infty} du u^{-b-1}\int_0^{2\pi} d\t
\int_0^{+\infty} dr r^{2a+1} e^{-u(1+r^2+2rcos\t)}=\cr
&={1\over \G(-b)} \int_0^{+\infty} du u^{-b-1} e^{-u}\int_0^{+\infty}
dr r^{2a+1} e^{-ur^2} \int_0^{2\pi}e^{-2urcos\t}=\cr
&={2\pi\over \G(-b)} \int_0^{+\infty} du u^{-b-1} e^{-u}\int_0^{+\infty}
dr r^{2a+1} e^{-ur^2} J_0(-2iur)=\cr}
$$
where $J_0(x)$ is the zero-order Bessel function of real argument, now
integrating in $r$ and $u$, we obtain
$$
\eqalign{
I(a,b)&=\pi {\G(a+1)\over \G(-b)} \int_0^{+\infty} du u^{-a-b-2} e^{-u}~_1F_1
(a+1,1,u)=\cr
&=\pi {\G(a+1)\G(-a-b-1)\over \G(-b)}~_2F_1 (-a-b-1,a+1,1,1)=\cr
&=\pi {\G(a+1)\G(b+1)\G(-a-b-1)\over \G(-a)\G(-b)\G(a+b+2)}\cr}
$$
that holds under the conditions
$$
Re~a>-1,~~Re~b>-1,~~Re(a+b)<-1
$$
which obviously comprehend also the conditions $b<0$ since
$$
a+b<-1,~~~a>-1~~\to~~b<-1-a<-1+1=0.
$$
\vglue 0.6cm
{\bf Appendix C \hfil}
\vglue 0.4cm
In this appendix we expose the calculation of the parameters involved
in the computation of three and four points correlator ($n=3$ and $n=4$)
for $B=1-1/d$ and arbitrary $d$, in particular the values
in the text are obtained by setting $d=4$. Substituting this $B$ in equations
of page $8$ we get
$$
\eqalign{
&Q_1(d)=2\sqrt{2} {\sqrt{2d^2+1-2d}\over d} -\left( 1-{1\over d}\right)
\sqrt{d\over 3}\cr
&Q_2 (d)= \sqrt{d\over 3}\cr
&\a(d)=-\sqrt{2} {\sqrt{2d^2+1-2d}\over d}\cr
&\beta(d,k)={\sqrt{2d^2+1-2d}\over d}(\sqrt{d}|k|-\sqrt{2})\cr
&a,b,p=\sqrt{2d}|k_{1,3,4}|-2\cr
&a',b'=\sqrt{2d} (|k_{1,3}|+|k_4|)-3|k_{1,3} k_4|-2\cr}
$$

Then by imposing condition \chargea\ with $s=1$ in the kinematic
$$
k_2<0,~~~~k_{i\not= 2}>0~~1\leq i\leq n
$$
we obtain an algebraic equation in the momenta
$$
\eqalign{
&{\sqrt{2d^2+1-2d}\over d}[\sqrt{d}(\sum_{i\not= 2} k_i -k_2)-n\sqrt{2}]-
\sqrt{2}{\sqrt{2d^2+1-2d}\over d}+2\sqrt{2}{\sqrt{2d^2+1-2d}\over d}+\cr
&-\left(1-{1\over d}\right) \sqrt{d\over 3}=0\cr
&\sum_{i\not= 2} k_i -k_2=(n-1)\sqrt{2\over d} +{d-1\over\sqrt{3}\sqrt{2d^2
+1-2d}}\cr}
$$

Considering now \chargeb\
$$
\sum k_i =0
$$
we can fix $k_2$
$$
k_2= {n-1\over 2}\sqrt{2\over d} +{d-1\over2\sqrt{3}\sqrt{2d^2
+1-2d}}
$$
and it remains just one relation between the other momenta.
\vglue 0.6cm
{\bf REFERENCES \hfil}
\vglue 0.4cm
\parindent=0.pt

{[1] A. M. Polyakov, Mod. Phys. Lett. A2 (1987) 893; V. G. Knizhnik, A. M.
Polyakov and A. B. Zamolodchikov, Mod. Phys. Lett. A3 (1988) 819}

{[2] J. Distler and H. Kawai, Nucl. Phys. B321 (1989) 509;
F. David, Mod Phys Lett. A3 (1988) 1651}

{[3] E. Brezin and V. Kazakov, Phys. Lett. 236B (1990) 144;
M. Douglas and S. Shenker, Nucl. Phys. B235 (1990) 635;
D. J. Gross and A. A. Migdal, Phys. Rev. Lett. 64 (1990) 717}

{[4] M. Cates, Europhys. Lett. 8 (1988) 719}

{[5] E. Witten, Phys Rev. D44 (1991) 314}

{[6] M. Martellini, M. Spreafico and K. Yoshida, Mod. Phys. Lett. A7 (1992)
1281}

{[7] A. B. Zamolodchikov, Phys. Lett. 117B (1982) 87}

{[8] T. Eguchi and H. Ooguri, Phys Lett. B187 (1987) 127}

{[9] J. Cohn and V. Periwal, Phys. Lett. B270 (1991) 18}

{[10] "Scaling Exponents in Quantum Gravity near Two Dimensions" H. Kawai,
Y. Kitazawa and M. Ninomiya, YITP preprint, YITP/U-92-05 (hep-th/9206081)}

{[11] V. A. Dotsenko and Vl. S. Fateev, Nucl. Phys. B240 (FS12)
(1984) 312; Nucl. Phys. B251 (FS13) (1985) 691}

{[12] A. Gupta, S. P. Trevedi and M. B. Wise, Nucl. Phys. B340 (1990) 475}

{[13] P. Di Francesco and D. Kutasov, Phys. Lett. 261B (1991) 385}

{[14] M. Martellini, M. Spreafico and K. Yoshida, Roma preprint 916 (1992)
to be published in the proceedings of International Workshop of String
Theory, 21/26-9-1992, Accademia dei Lincei, Roma}

{[15] H. Bateman, "Higher Transcendental Functions", McGraw-Hill, 1953}

{[16] J. Ambjorn, B. Durhuus, J. Fr$\ddot o$hlich and P. Orland, Nuc.
Phys. B270 (1986) 457}

{[17] M. Caselle, F. Gliozzi and S. Vinti, DFTT preprint, DFTT-12/93}

{[18] L. Baulieu, C. Becchi and R. Stora, Phys. Lett. 180B (1986) 55}

{[19] L. Baulieu and M. Bellon, Phys. Lett. 196B (1987) 142}

{[20] C. Becchi, Nuc. Phys. B304 (1988) 513}

{[21] R. Stora in Carg\`ese Summer Study Institute "Non Perturbative Quantum
Field Theory" (1989)}

{[22] L. Baulieu  in Carg\`ese Summer Study Institute "Non Perturbative Quantum
Field Theory" (1989)}

{[23] L. Bonora, M. Martellini and C. M. Viallet, Phys. Lett. 197B (1987) 335}

{[24] K. Fujikawa, Nucl. Phys. B291 (1987) 583}

{[25] K. Fujikawa, T. Inagaki and H. Suzuki, Phys. Lett. 213B (1988) 279}

{[26] K. Yoshida, Mod. Phys. Lett. A4 (1989) 71}

{[27] M. N. Sanielvici, G. W. Senenoff and Y. Shi-Wu, Phys. Rev. Lett. 60
(1988)
2571}

{[28] H. Kawai and R. Nakayama, "Quantum $R^2$ Gravity in Two Dimension"
KEK Preprint 92-212 (1993)}

{[29] For general ideas of "Effective action" see S. Weimberg
"Phenomenological Lagrangians" Physica 96A (1979) 327 and "Conference Summary"
of XXVI International High Energy Physics, Dallas, Texas, Aug. 1992; for formal
definition G. Jona-Lasinio, Nuovo Cimento 34 (1964) 1790, B. Zumino in
"Lectures
in Elementary Particles and Quantum Field Theory" ed. S. Deser (MIT, Cambrige,
Mass., 1970)}
\vfill
\endpage
\figout
\vfill
\endpage
\fig{Behaviour of $\G(d)$ for arbitrary values of $d$.}
\end